\documentstyle[aps,prd,preprint,tighten]{revtex}
\begin{document}
\title{Phenomenological quark model for baryon magnetic
moments and beta decay ratios ($G_A/G_V$)}
\author{Jerrold Franklin\footnote{Internet address:
V1357@ASTRO.TEMPLE.EDU}}
\address{Department of Physics\\
Temple University Philadelphia, PA 19122-6082}
\date{March 20, 2001}
\maketitle
\begin{abstract}

Baryon magnetic moments and beta decay ratios ($G_A/G_V$)
are calculated in a phenomenological quark model.   
Non-static effects of  pion exchange and orbital excitation are included.  
Good agreement with experiment is found for a combined fit to all measured baryon magnetic moments and beta decay ratios.  The model predicts an antiquark content for the proton that is consistent with the Gottfried sum rule.
\end{abstract}
\section{INTRODUCTION}

The original static quark model (SQM) made predictions for baryon magnetic
moments[1-3] that were in remarkable qualitative agreement with early
magnetic moment measurements.  However, more accurate measurements of the
magnetic moments of the baryon octet differ from the SQM predictions by up to 0.2 nuclear magnetons.  Also, the SQM can not be reconciled with the ratio $G_A/G_V$ of beta decay constants in baryon beta decay.

These quantitative failures of the SQM
have generally been attributed to various ``non-static" effects
in the quark model.  These non-static effects must break SU(3) 
symmetry if they are to improve the agreement of magnetic moment 
predictions with experiment.
This can be seen from the disagreement with experiment of the sum rules\cite{jf69}
\begin{equation}
   \mu(p) -\mu(n) +\mu(\Sigma^-) -
   \mu(\Sigma^+) + \mu(\Xi^0) - \mu(\Xi^-)=0 \quad  (0.49\pm .05)
\label{eq:srule1}
\end{equation}
and
\begin{equation}
   \mu(p)+2\mu(n) +\mu(\Xi^-) - \mu(\Xi^0)=0 \quad  (-0.43\pm .01).
\label{eq:srule2}
\end{equation}
The most recent experimental
value\cite{pdg} for each sum rule is shown in parentheses in 
Eqs.\ (1) and (2).

For the baryon combinations in each sum rule, the non-static magnetic moment contributions would cancel if the ultimate contribution from each quark were independent of which baryon the quark was in.  This ``baryon independence" would follow, for instance, if the non-static parts of the baryon wave functions were SU(3) symmetric.  Because of the cancellation of the non-static contributions, it
was originally expected that the sum rules would be in better
agreement with experiment than individual quark moments.  However, subsequent
tests of the sum rules showed that they disagreed with experiment by
more than did any single magnetic moment\cite{jfk}.  The violation of the sum rules indicates 
that strong SU(3) breaking and baryon dependent non-static contributions are
required for baryon magnetic moments.

The admixture of pion configurations  to the quark  model wave 
functions has been proposed\cite{jfpi} as an important SU(3) breaking non-static 
effect that would break the sum rules 
of Eqs.\ (1) and (2).  Such pion contributions were shown in I to  improve quark
model magnetic moment predictions significantly.  But there was still substantial disagreement with experiment for some of the moments. 

In this paper we show that the inclusion of orbital excitation, along with the pion contribution, permits us to extend the model to simultaneously fit magnetic moments and the beta decay ratios $G_A/G_V$, along with a better overall agreement with experiment.  
It had been very difficult to reconcile the quark model magnetic moment predictions with quark model beta decay ratios, especially $G_A/G_V$ for neutron decay. 
The combination of the non-static effects (pionic and orbital) now
makes it possible with the same quark model to achieve good agreement with experiment for 
the combined set of baryon magnetic moments and beta decay constants.

In Section 2 of this paper, we review the phenomenological treatment in I of pion
components in the baryon wave functions, and the effect of pions on baryon magnetic
moments.  We extend the effect of pion admixtures to baryon beta decay as well.  
Section 3 adds an orbital component to the three quark wave function that leads to an orbital contribution to the magnetic moments and beta decay ratios.
In Section 4, we discuss $\Lambda$-$\Sigma^0$ mixing which should be included in any calculation of this nature.  In Section 5 the three effects (pionic, orbital contribution, and mixing) are combined to achieve a good fit of all measured baryon moments and beta decay $G_A/G_V$ ratios.  
In Section 6, the model parameters are used to determine the quark spin distributions of the proton in its rest frame.  We also calculate an antiquark content of the proton that is consistent with the Gottfried sum rule.  
We state our major conclusions in Section 7.
\section{PION CONTRIBUTIONS TO BARYON MAGNETIC MOMENTS
AND $G_A/G_V$} 

A detailed calculation of pion contributions to baryon magnetic moments is 
given in I.  In this section we review that calculation, and extend it to the
ratio $G_A/G_V$ for baryon beta decay.  
There are two kinds of pion contribution.  If pions are created and then absorbed by the same quark, they affect only that quark's anomalous moment.
This contribution is independent of which quark the baryon is in.  This means it cannot affect the magnetic moment sums in Eqs.\ (1) and (2), and so cannot improve the overall prediction  for baryon magnetic moments.   The absorption of an emitted charged pion by a different quark in the same baryon leads to exchange currents.  These are different for different baryons.
For instance, the u quark in a proton can emit a positive pion that is then absorbed by the d quark in the proton.  But this type of exchange current cannot occur in a $\Sigma^+$ hyperon where there is no d quark.  Because these pion exchange contributions are baryon dependent, they  do affect the sum rules, and can improve the prediction of baryon magnetic moments\cite{jfc}.

If the exchange currents were SU(3) symmetric, then kaon and eta exchange currents would compensate for the pion exchange currents, preserving the disagreement with experiment of the sum rules.  In I, and here, we assume that pion exchange dominates because of the 
particularly small mass of the pion.  The effect of the heavier meson 
exchanges is neglected, breaking SU(3) as is necessary to improve
agreement with experiment. 

Any theory with full conservation of isotopic spin without SU(3) symmetry,
such as occurs when pions dominate the exchange, will include
baryon dependent charge exchange magnetic moment contributions.
In our phenomenological representation of the pion component of baryon 
wave functions, isotopic spin is conserved at both the quark and the baryon level.  
This provides the proper mix of direct and exchange pion currents without specifying any
specific mechanism of pion emission.  The procedure for this is shown in
detail in Section 3 of I.  Here we outline the steps followed in I.

Physical baryon states $|B>$ for each isomultiplet, including pionic parts, are defined by
\begin{eqnarray}
 |N>&=&\alpha_N\, N + \beta_N [N\pi]+\delta_N[\Delta\pi]\\
|\Sigma>&=&\alpha_{\Sigma}\, \Sigma + \beta_{\Sigma} [\Sigma\pi]
+\delta_{\Sigma}[\Sigma^*\pi]+\epsilon_{\Sigma} [\Lambda\pi]\\
 |\Xi>&=&\alpha_{\Xi}\, \Xi + \beta_{\Xi} [\Xi\pi]+\delta_{\Xi}[\Xi^*\pi]\\
 |\Lambda>&=&\alpha_{\Lambda}\, \Lambda 
+ \beta_{\Lambda}[\Sigma\pi]+\delta_{\Lambda}[\Sigma^*\pi]\\
 |\Delta>&=&\alpha_{\Delta}\,\Delta + \beta_{\Delta} [N\pi]+
\delta_{\Delta}[\Delta\pi]\\
|\Omega>&=&\Omega^-.
\end{eqnarray}
In each of Eqs. (3)-(8), the notation $B(=N$, $\Sigma$, $\Xi$, $\Lambda$, $\Delta$, 
$\Omega^-)$ represents the static quark model wave function of each baryon, and
$[B\pi]$ represents the appropriate linear combinations
of isotopic spin and angular momentum states of the static quark model baryons $B$ and pions (with L=1 for the pions).  An explicit example 
for the proton is given by Eq. (5) of I.  The $\Omega^-$ baryon cannot emit pions, and so the physical $|\Omega>$ is the same as the quark model $\Omega^-$.

The expansion coefficients $\beta$, $\delta$, $\epsilon$ for each baryon are determined by matrix elements of a general pion emission operator
\begin{equation}
\Theta_{\pi} =
\gamma\sum_{i=1}^3\mbox{\boldmath $\sigma$}^i\mbox{\boldmath $\cdot{\hat p}$}_i
\mbox{\boldmath $\tau$}^i\mbox{\boldmath $\cdot\phi$}_\pi.
\end{equation}
between quark model states.  The coefficient $\alpha$ is then determined from the normalization 
of the physical baryon state.
For any quark model state B, the corresponding 
physical baryon state is
\begin{equation}
|B>=(\alpha +\Theta_{\pi})B.
\label{eq:phys2}
\end{equation}
In I, the physical baryon states produced by the pion emission operator of Eq.\ (\ref{eq:phys2})
are compared to the physical baryon states in Eqs.\ (3)-(8) to determine the 
expansion coefficients  $\beta$, $\delta$, $\epsilon$ in terms of 
the pion emission coefficient $\gamma$.  These expansion coefficents are 
listed in Table 1.  
\begin{table}
\begin{center}
\caption{Expansion coefficients for physical baryon states.}
\begin{tabular}{cccc}
Baryon & $\beta$ & $\delta$ & $\epsilon$ \\
\hline
N &$ 5\gamma$ & $4\sqrt{2}\gamma'$ & - \\
$\Sigma$ &$ \sqrt{32/3}\gamma$ & $\sqrt{16/3}\gamma'$ &$ 2\gamma$ \\
$\Xi$ & $-\gamma$ & $2\sqrt{2}\gamma'$ & - \\
$\Lambda$ &$-2\sqrt{3}\gamma'$ & $2\sqrt{6}\gamma'$ & - \\
$\Delta$ & $2\sqrt{2}\gamma$ & $5\gamma$ & - \\
\end{tabular}
\label{tab:coeffs}
\end{center}
\end{table}

In Table 1, we have distinguished between $\gamma$, the pion emission coefficient for octet baryon states, and $\gamma'$ the pion emission coefficient connecting octet states to decuplet states.  Because of the higher masses of the decuplet states, the energy denominators would be larger and the overlap integrals smaller for the octet-decuplet transition than for octet-octet transitions in any calculation of  $\gamma$.  So we should expect that $\gamma'$ will be smaller than $\gamma$.  The $\Lambda$ is considerably lighter than the $\Sigma$, and its wave function is different than that of the $\Sigma$, so we also use $\gamma'$ for the $\Lambda$.

The magnetic moment operator including the pion contribution is
\begin{equation}
\mbox{\boldmath $\mu$}_{op}
=\sum_{i=1}^3\mbox{\boldmath $\sigma$}^i\mu_i+{\bf L}_{\pi}M_L,
\end{equation}
where the $\mu_i$ are the quark Dirac moments.   We use the Dirac moment for the quarks, because we assume that any anomalous moment of the quarks is produced by the pion contributions
of this paper.  The pion is emitted in an $L=1$ state and
there is an effective orbital moment $M_L$. 
We apportion the orbital moment between the pion and the recoiling baryon according to the center of mass relations
\begin{eqnarray}
M_L &=& M_\pi+M_B\\
M_\pi & = & \frac{e_\pi M}{\left[1+\frac{m_p}{m_B M}\right]}\\
M_B & = & \frac{(e_B m_p/m_B)}{\left[1+\frac{m_B M}{m_p}\right]}.
\end{eqnarray}
The orbital moments $M_\pi$ and $M_B$ are given in nuclear magnetons.  The charges
$e_\pi$ and $e_B$ are $\pm 1$, depending on the charge of the particular particle. 
The orbital moments depend on the masses, $m_B$ and $m_p$ and the ratio of the proton mass to
an effective pion mass
\begin{equation}
M=m_p/m_\pi({\rm effective}).
\end{equation}
Since the pion motion is, in fact, relativistic we take the ratio $M$ to be an adjustable parameter in fitting the baryon magnetic moments.

The baryon magnetic moments 
are given by the expectation value of  $\mbox{\boldmath $\mu$}_{op}$ in the physical
baryon states given by Eqs.\ (3)-(8).
The calculation leads to Eqs.\ (A1)-(A8) for eight
octet baryon magnetic moments in the Appendix of I.  We reproduce these
equations here\cite{sl}, along with additional results for 
the decuplet moments $\mu_{\Omega^-}$ and $\mu_{\Delta^{++}}$, and for the transition moment $\mu_{\Delta p}$ . 
\begin{eqnarray}
\mu_p &  = & p+50g(-5p-n+2M_\pi+M_p)\nonumber\\
& & +32g'(-9p-20d+M_\pi-4M_\Delta)-640d\sqrt{gg'}
\label{eq:momp}\\
\mu_n & = &  n+50g(-5n-p-2M_\pi+2M_p)\nonumber\\
& & +32g'(-9n+5d-M_\pi+M_\Delta)+640d\sqrt{gg'}\\
\mu_{\Sigma^+}&  = &\Sigma^+ +4g[-37\Sigma^+ -4\Sigma^0 -3\Lambda + 4\sqrt{3}(\Sigma,\Lambda)+14M_\pi+8M_\Sigma] \nonumber\\
& &  +8g'[-6\Sigma^+ +\frac{5}{3}(-5d+2s)
-M_\pi-M_{\Sigma^*}]-\frac{16}{3}(38d+16s)\sqrt{gg'}\hspace*{1cm}\\
\mu_{\Sigma^-}&  =& \Sigma^- +4g[-37\Sigma^- -4\Sigma^0 -3\Lambda 
-4\sqrt{3}(\Sigma,\Lambda)-14M_\pi-8M_\Sigma]\nonumber  \\
&  &+8g'[-6\Sigma^+ +\frac{5}{3}(d+2s)+M_\pi+M_\Sigma*]
+\frac{16}{3}(22d-16s)\sqrt{gg'}\\
\mu_{\Xi^0} & =& \Xi^0 +2g(-5\Xi^0 -\Xi^- +2M_\pi-2M_\Xi)\nonumber\\
& & +8g'(-9\Xi^0+10s-2M_\pi+2M_\Xi)+32s\sqrt{gg'}\\
\mu_{\Xi^-}& =& \Xi^- +2g(-5\Xi^- -\Xi^0 -2M_\pi-M_\Xi)\nonumber\\
& & +8g'(-9\Xi^- -5d+10s+2M_\pi+M_\Xi)+32(s+d)\sqrt{gg'}\\
\mu_{\Lambda} & = & \Lambda+12g(-9\Lambda-\Sigma^+
-\Sigma^0-\Sigma^-)\nonumber\\
& & +24g'(-9\Lambda-5d+5s)+96(d+2s)\sqrt{gg'}
\label{eq:moml}\\
\mu_{\Sigma\Lambda}& = & [\alpha_\Sigma\alpha_\Lambda-12\sqrt{gg'}] (\Sigma,\Lambda)
+16\sqrt{3}g'(-3d+M_\pi-M_{\Sigma^*})\nonumber\\
& & + 8\sqrt{3}\sqrt{gg'}(-14d +\Sigma^+ -\Sigma^- +4M_\pi -4M_\Sigma)\\
\mu_{\Omega^-} & = & \Omega^-\\
\mu_{\Delta^{++}} & = & \Delta^{++}+6g(12p+125d+18M_\pi+12M_p+24M_\Delta)\\
\mu_{\Delta p} & = &-2\sqrt{2}[\alpha_p\alpha_\Delta d
-5g(4p-4n-5d+2M_\pi-2M_p)\nonumber\\
& & -2\sqrt{gg'}(-54d+25M_\pi-25M_\Delta)] 
\label{eq:momdp}
\end{eqnarray}

On the right hand sides of Eqs.\ (16)-(26), baryon symbols have been used 
to represent static quark model magnetic moments, while quark symbols represent quark magnetic moments coming from baryon resonances in the quark model.
We have replaced the $u$-quark moment by using $u=-2d$, corresponding
to our use of Dirac moments for the quarks.
The constants $g$ and $g'$ are related to the pion emission constants by
\begin{equation}
g=\gamma^2/9\quad {\rm and}\quad  g'=\gamma'^{2}/9.
\end{equation}

The pion admixtures in the physical baryon states affect 
magnetic moments in three ways:
\begin{enumerate}
\item An orbital magnetic moment due to the fact that the pions are emitted in an $L=1$ state.
\item  The quark model magnetic moments for the recoil baryons.
\item   The decrease in the bare baryon probability given 
by the normalization condition
\begin{equation}
\alpha^2=1-\beta^2-\delta^2-\epsilon^2.
\end{equation}
\end{enumerate}

We see from the above derivation that the pion contribution to baryon magnetic moments depends on three parameters.  We take these to be
\begin{enumerate}
\item  The probability that the physical nucleon contains one pion 
\begin{equation}
P_\pi = 9(25g+32g'),
\end{equation}
\item  The proton/pion effective mass ratio $M$.
\item  The ratio of $\Delta$-$\pi$ to $N$-$\pi$ probabilities in the nucleon
\begin{equation}
R_\Delta = \frac{32g'}{25g}.
\end{equation}
Note that this definition of $R_\Delta$ differs from the ratio $R$ 
given in I by the factor (32/25). 
\end{enumerate}

The beta decay constants are given by matrix elements of the operators
\begin{equation}
{\hat G}_A=\sum^3_i\sigma^i_z(\tau^i_+{\rm\, or\; }v^i_+) 
\label{ga}
\end{equation} 
and
\begin{equation}
{\hat G}_V=\sum^3_i(\tau^i_+{\rm\, or\; }v^i_+),
\end{equation}
where $\tau_+$ and $v_+$ are isotopic spin (used for neutron beta decay) and $v$-spin (used for hyperon beta decay) raising operators.

${\hat G}_A$ has no explicit pion part because the  pion with $J^P=0^-$ does not contribute directly to $G_A$.  ${\hat G}_V$ would have a pion part.  We do not include it because we need apply ${\hat G}_V$ to only the quark model wave function to get $G_V$ for any baryon.  Then, because the vector current is conserved (CVC),  including pions will not change $G_V$.

Using the operators ${\hat G}_A$ and ${\hat G}_V$ between the physical baryon states of
Eqs.\ (3)-(8) gives, after some algebra, the following results for the ratios
 $G_{A/V}=G_A/G_V$
\begin{eqnarray}
G_{A/V}(n\rightarrow p) & = & \frac{5}{3}[1-200g-128g'+256\sqrt{gg'}]
\label{eq:gpin}\\
G_{A/V}({\Lambda}\rightarrow p) & = & 1[\alpha_N\alpha_{\Lambda}
+288g'+90\sqrt{gg'}]\\
G_{A/V}(\Xi^-\rightarrow\Lambda)) & = & \frac{1}{3}
[\alpha_{\Lambda}\alpha_{\Xi}-30g+240g'-48\sqrt{gg'}]\\
G_{A/V}(\Sigma^-\rightarrow n) & = & -\frac{1}{3}[\alpha_N\alpha_\Sigma 
+50g-160g'+416\sqrt{gg'}]\\
G_{A/V}(\Xi^0\rightarrow\Sigma^+)) & = &\frac{5}{3}
[\alpha_{\Sigma}\alpha_{\Xi}+34g/5 - 32g' + 192\sqrt{gg'}/5]
\label{eq:gpix}\\
G_{A/V}(\Delta^+\rightarrow p) & = & -2\sqrt{\frac{2}{3}}
[\alpha_p\alpha_\Delta + 450g + 216\sqrt{gg'}].
\label{eq:gad}
\end{eqnarray}
The normalization constants $\alpha_B$ are given by
\begin{eqnarray}
\alpha_N & = & \sqrt{1-225g-288g'}
\label{eq:an}\\
\alpha_\Lambda & = & \sqrt{1-324g'}
\label{eq:al}\\
\alpha_\Xi & = & \sqrt{1-9g-72g'}\\
\alpha_\Sigma & = & \sqrt{1-132g- 48g'}\\
\alpha_\Delta & = & \sqrt{1-297g}.
\label{eq:ad}
\end{eqnarray}

\section{Orbital Excitation}

There are seven different types of orbital excitation that could affect the magnetic moment of a three quark bound state of spin $\frac{1}{2}$.  These are listed in the Appendix of  Ref.\ [3].  
(At that early stage of the quark model, the possibility of a ground state with orbital angular momentum had not been ruled out, and this resulted in the angular momentum $l_0$ appearing in the angular states.  Now it is known that the $l_0$ appearing there is zero, so that it can just be left out of the equations.)  

Of these angular momentum states, we expect that the state with Dalitz angular momenta $l=1$, $L=1$, with $L+l=0$ to be the most important.  This state is listed as state (5) in Ref.\ [3].  
It is sometimes referred to as the $S'$ state because it has no total orbital angular momentum.
States (1)-(4$\pm$) in Ref.\ [3] have $l$ or $L$ of 2, and some must couple to  total quark spin of  $\frac{3}{2}$.  State (6) also has $l=L=1$, and could be of comparable size with the $l+L=0$ state,
but turns out not to have as much effect on magnetic moments and beta decay ratios.\cite{rcp}

The effect of the $l+L=0$ orbital state on baryon magnetic moments is through the change in the quark spin states.  Two identical quarks must now be in a spin zero state, so that the magnetic moment of the six baryons with two identical quarks will be that of the odd quark.  Taking into account the decrease in the normalization of the ground state, the change in the baryon magnetic moment  for these six baryons is
\begin{equation}
\Delta\mu_B = \eta(\mu_{q'}-B),
\end{equation}     
where $\mu_{q'}$ is the magnetic moment of the unlike quark,
and $B$ represents the static quark model magnetic moment of baryon $B$.
The coefficient $\eta$ is the probability for the physical baryon to be in the $l+L=0$ state.
The quark spin states of the $\Lambda$ and $\Sigma^0$ are just interchanged in the $l+L=0$ state.
This results in
\begin{equation}
\Delta\mu_{\Lambda} = \eta(\Sigma^0 - \Lambda).
\end{equation}
These orbital additions should be added to the baryon magnetic moments listed in Eqs.\ (\ref{eq:momp})-(\ref{eq:moml}) 

The $l+L=0$ excitation also affects the beta decay ratios.  These can be calculated from the spin states of the baryons for this state, and an angular overlap integral over the internal cordinates.
For this purpose, it helps to write the $l+L=0$ state in terms of the two vectors, $\bf r$ and 
\mbox{\boldmath $\rho$}.  $\bf r$ is the vector between the two like quarks in a baryon, and 
\mbox{\boldmath $\rho$} is the vector from the midpoint of $\bf r$ to the third, unlike quark.
Then the wave function for this state can be written as
\begin{equation}
\Psi(\mbox{\boldmath $r,\rho$}) = \mbox{\boldmath $r\cdot\rho$}\psi_0(r,\rho)
\chi',
\end{equation}
where $\psi_0(r,\rho)$ is spherically symmetric in both vectors, and $\chi'$ is the spin state
\begin{equation}
\chi' = \frac{1}{\sqrt{2}}(\uparrow\downarrow\uparrow-\downarrow\uparrow\uparrow).
\end{equation}

For a beta decay like $n\rightarrow p$, the matrix element to be evaluated is
\begin{equation}
\Delta G_A(n\rightarrow p) = <\Psi(-uud)|{\hat G_A}|\Psi(udd)>.
\label{dg}
\end{equation}
Note that in the two wave functions, the quark ordering is permuted for the two like quarks.
This leads to a factor of $-\frac{1}{2}$ from the angular integral for these wave functions.
This factor of $-\frac{1}{2}$ enters for all beta decay matrix elements except for the $\Sigma^-$ to neutron decay where the overlap factor is +1.  The spin projections for the beta decay operator
${\hat G}_A$ are then straightforward, and the results are
\begin{eqnarray}
\Delta g_{A/V}(n\rightarrow p) & = & -\frac{1}{2}\eta\\
\Delta g_{A/V}(\Lambda\rightarrow p) & = & -\frac{1}{6}\eta\\
\Delta g_{A/V}(\Xi^-\rightarrow\Lambda) & = & -\frac{1}{6}\eta\\
\Delta g_{A/V}(\Sigma^-\rightarrow n) & = & +\eta\\
\Delta g_{A/V}(\Xi^0\rightarrow \Sigma^-) & = & +\frac{1}{2}\eta.
\end{eqnarray}
These orbital corrections should be added to the beta decay ratios given in 
Eqs.\ (\ref{eq:gpin})-(\ref{eq:gpix}).  The coefficient -$\eta$ should also be included in the square roots for the normalization constants in Eqs.\  (\ref{eq:an})-(\ref{eq:ad}).

\section{$\Lambda-\Sigma^0$ mixing.}

It has been known for some time that quark model predictions should be corrected for mixing of the $\Lambda$ and $\Sigma^0$ quark states \cite{msdv,jf68,isg,flnc,jfpi,karl,jfx}.  
This mixing is a necessary result of a spin dependent off-diagonal matrix element connecting the  
$\Lambda$ and $\Sigma^0$ that is inherent in any quark model.  This mixing should be included in any consistent quark model calculation at this level of accuracy, but is often left out.  

The mixing formalism is given in detail in Refs.\ \cite{flnc} and \cite{jfx}.  Here we list the relevant formulae for this paper.  We are using a different sign convention for the $\Sigma$ quark model wave function here than previously, so that some of the signs are different.  We also consistently make small angle approximations here for the mixing angle $\theta$.
The physical $|\Sigma^0$$>$ and $|\Lambda$$>$ have mixtures of the other hyperon, $\Lambda$ and $\Sigma^0$, given by
\begin{eqnarray}
|\Sigma^0\!\!> & = & \Sigma^0 - \theta\Lambda\\
|\Lambda\!> & == & \Lambda  +\theta\Sigma^0
\end {eqnarray}

Although the mixing angle probably comes from a combination of magnetic and QCD interactions,  the formalism in Ref.\ \cite{flnc} applies for any mixing mechanism in the quark model. 
The off-diagonal matrix element connecting the $\Sigma^0$ and $\Lambda$ can be related directly
to a combination of hyperon mass differences that give the result
\begin{equation} 
\theta =\frac{m_{\Sigma^{*-}}-m_{\Sigma^{*+}}-m_{\Sigma^-}+m_{\Sigma^+}}
{2\sqrt{3}(m_{\Sigma^0} - m_{\Lambda})}=-0.014\pm .004\; \rm radians.
\end{equation}

The mixing leads to the following small additions to quark model magnetic moments and beta decay constants involving the $\Lambda$ or $\Sigma^0$:
\begin{eqnarray}
\Delta\mu_\Lambda & = &  + 2\theta\; \mu_{\Sigma\Lambda} = -0.045\pm .013\; \rm nm
\label{mix1}\\
\Delta\mu_{\Sigma\Lambda} & = &  
+ \theta\; (\mu_{\Sigma^0}-\mu_\Lambda) =-0.03\pm .01\; \rm nm\\
\Delta G_{A/V}(\Lambda\rightarrow p) & = &-\frac{4}{3\sqrt{3}}\;\theta=-0.01\\
\Delta G_{A/V}(\Xi^- \rightarrow \Lambda) & = &+\frac{4}{3\sqrt{3}}\;\theta=+0.01.
\label{mix4}
\end{eqnarray}

\section{Results.}

In this section we provide the results of a $\chi^2$ fit to experiment of ten magnetic moment predictions and five beta decay ratio predictions.  
The model predicts quark model magnetic moments and beta decay constants,
 modified by pion direct and exchange currents, and orbital excitation.  The static quark model involves two parameters, the input masses of the nucleon and strange quarks.  The pion contribution involves three additional parameters, $P_\pi$, the percentage of pion admixture in the nucleon,  $M$, the effective pion magnetic moment,
and $R_\Delta$,  the ratio of $\Delta$-$\pi$ to $N$-$\pi$ admixture in the nucleon.  
The orbital contribution is characterized by the probability $\eta$ of the orbital excitation.
So that we are fitting fifteen experimental quantities with six parameters, corresponding to nine degrees of freedom (DF). 

The results of this fit are shown in table 3.  The pure quark model two parameter fit,
and the fit with only the pion contribution are also shown for comparison.
We have also included the model prediction for the beta decay ratio $G_{A/V}(\Delta^{++}$$\rightarrow$$p)$, which is used in the calculation of weak proton capture on $^3He$\cite{mar}.  The resonance transition moment $\mu(\Delta^+ p)$ is not included in the fit because its experimental determination is not clear.   
All the magnetic moments are in units of nuclear magnetons (nm), while the beta decay ratios are pure numbers.  In determining $\chi^2$, we have used a theoretical error of 0.05 for $G_{A/V}$ and
0.05 nm for the magnetic moments, added in quadrature with the experimental errors.  This is used to avoid having the fit to experiment arbitrarily dominated by the most accurate measurements.
Also, there are a number of small effects that are expected to be of this order that have been left out of the calculation.     
\newpage
\begin{table}
\begin{center}
\caption{Fit of the quark model with pion and orbital contributions.  Experimental values are from Ref.\ [5], except where noted otherwise.}
\begin{tabular}{llrrr}
 & \hspace*{.2cm}Expt. & SQM\hspace{.5cm} & Pion\hspace{.5cm} & Pi+Orbital \\
\hline
$\mu(p)$	& \hspace{.128cm}2.79 & 2.75$\;\,$(0.7)&2.65 (7.7)   & 2.68 (5.1)\\
$\mu(n)$	&  -1.91 & -1.84$\;\,$(1.9)& -2.04(6.7) & -1.99 (2.3)\\
$\mu(\Sigma^+)$ & \hspace{.128cm}2.46$\pm$.01 &2.65 (14.7)&2.53(2.0) &  2.52 (1.5)\\
$\mu(\Sigma^-)$ &  -1.16$\pm$.03 &-1.02$\;\,$(6.7)&-1.14 (0.2) &  -1.17 (0.0)\\
$\mu(\Xi^0)$ &  -1.25$\pm$.01 & -1.44(13.7)&-1.42(10.7) &  -1.27 (0.2)\\
$\mu(\Xi^-)$ &  -0.65$\pm$.00 & -0.52$\;\,$(6.3)&-0.54 (4.8) &  -0.59 (1.6)\\
$\mu(\Lambda)$ &  -0.61$\pm$.00 & -0.67$\;\,$(1.2)& -0.67 (1.1) &  -0.56 (1.0)\\
$\mu(\Sigma,\Lambda)$ &\hspace{.06cm}  1.61$\pm$.08 & 1.57$\;\,$(0.2)&1.46 (2.6)
 &  1.51 (1.0)\\
$\mu(\Omega^-)$ &  -2.02$\pm$.05 & -1.87$\;\,$(4.6)&-1.91 (2.2) &  -2.07 (0.5)\\
$\mu(\Delta^{++})$\cite{mud} &\hspace*{.05cm} 6.22$\pm$.7 & 5.50 (1.8)
 &5.49 (1.9)& 6.17 (0.0)\\
$\mu(\Delta^+$,p) & &  2.59\hspace*{.95cm}&2.49\hspace{.95cm}
& 2.79\hspace*{.95cm}\\
\hline
$G_{A/V}$(n,p)& \hspace{.128cm}1.27$\pm$.00 & 1.67$\;\;$(64) &1.33 (1.8)  &1.32 (1.3)\\
$G_{A/V}(\Lambda$,p)& \hspace{.128cm}0.72$\pm$.02 & 1.00$\;$ (27) &0.86 (6.9) &0.78 (1.6)\\
$G_{A/V}(\Xi^-,\Lambda$)& \hspace{.128cm}0.25$\pm$.05 &  0.33 (1.9)&0.30 (0.6) &0.24 (0.0)\\
$G_{A/V}(\Sigma^-$,n)& -0.34$\pm$.02 & -0.33$\;$(0.0) &-0.30 (0.4)& -0.21 (6.2)\\
$G_{A/V}(\Xi^0,\Sigma^+$)\cite{xz}& \hspace{.128cm}1.24$\pm$.27 &  1.67 (6.0)
& 1.53 (1.1)& 1.38 (0.3)\\
$G_{A/V}(\Delta^{++}$,p) & & -1.63\hspace{.95cm} &-2.09\hspace{.95cm}
 & -2.08\hspace{.07cm}$\pm .06$\\
\hline
$\chi^2-DF$ &  & $52-8$ & $51-10$  &$23-9$\\
\hline
$m_u$ (MeV) & &340 &340 & $297\pm 20$\hspace*{.13cm}\\
$m_s$ (MeV) & & 500 &490  &$453\pm 20$\hspace*{.13cm}\\
$P_\pi$ & & 0 &29\% & $33\pm 7\%$\\
$M(\pi)$ (nm) & &  &4.8& $4.7\pm 1.0$\\
$R_\Delta$ & & &3\% &$8\pm 5\%$\\
$\eta$(orbital) & & & 0&$8\pm 2\%$\\
\end{tabular}
\label{tab:moms}
\end{center}
\end{table}

The $\chi^2$ fit for the static quark model (SQM) in Table 3 does not include the beta decay ratios.  It is clear that the SQM is especially bad for neutron decay, and including it would raise $\chi^2$ to well over 100.  Among the magnetic moments, the Sigmas and the Xis are the worst fit for the SQM.  Including Pion exchange considerably improves the magnetic moment fits.  The Sigma problem is corrected, but there is still a mismatch between the Xi and the nucleon moments.
The most remarkable feature of the pion fit is the great improvement in $G_{A/V}$ for the neutron.
This permits an overall fit to both beta decay ratios and magnetic moments.
But this still is not enough to achieve really good agreement with experiment.  
Finally, adding the orbital state is seen to achieve a reasonable fit.

The best fit parameters for the (pi+orbital) case are shown at the bottom of table 3.
The $\pm$ values on the parameters  correspond to an increase in $\chi^2$ of $\chi^2/DF$.  The parameters all have reasonable values.  The probability of pions in the physical nucleon is rather high, but $M$ is close to the orbital magnetic moment for a pion of the physical mass.  
Although $R_\Delta$ is not large, the decuplet cannot be completely left out.  
Doing so increases $\chi^2$ to 35.  

The importance of each effect can be judged by the effect on $\chi^2$ when it is left out.
Leaving out the orbital excitation ($\eta$=0) increases $\chi^2$ to 51, while leaving out the pion exchange ($P_\pi$=0) increases $\chi^2$ to 104.  So it is clear that a combination of non-static effects (in this model, pion exchange, decuplet baryons, and orbital excitation)
is required to achieve a reasonable fit to all baryon moments and beta decay ratios.
That is why so many earlier calculations that concentrated on only one non-static effect could not achieve good overall fits.

The $\Lambda$-$\Sigma$ mixing, discussed in Section 4  
has been included in all four $\Lambda$ entries shown in table 3.  The mixing is a barely measurable effect in this fit.  $\chi^2$ increases by 2 if mixing is left out, almost all of the increase coming from a slightly worse prediction for the $\Lambda$ magnetic moment without mixing.

If the theoretical error of 0.05 is raised to 0.08, then $\chi^2$ is 9, and equal to the number of degrees of freedom.  So that 0.08 could be considered the level of accuracy of the model when fitting to this data.  Actually, to bring $\chi^2$ down to the number of degrees of freedom for 
the 0.05 theoretical error
would require shifts in some predictions of much less than 0.05.  There are a number of small effects we have left out that could be close enough to 0.05 to improve the accuracy.
Also, all six parameters which have been kept constant  should vary a bit from baryon to baryon, which could considerably reduce $\chi^2$.  To achieve this would require an accurate detailed calculation in a specific theory, beyond the simple phenomenological model considered here.

At this point, it is important to discuss the actual experimental significance of the $G_{A/V}$ ratios listed in Ref.\ [5].  As the discussion on page 694 of Ref.\ [5] indicates, the listed ratios are actually theoretical numbers, derived from experiment using the SU(3) symmetry assumption that the coupling parameter $g_2$ is zero.  This assumption is not confirmed by experiment, and SU(3) symmetry is at sharp variance with the model used here.  Consequently, it is of interest to see the effect on our fit of letting $g_2$ vary freely in interpreting the experimental distributions.  
This has been done in only one experiment, the study of the $\Xi^-$$\rightarrow$$n$ decay by 
Hsueh {\it et al.}\cite{hsueh}.  The experimental result is
$G_{A/V}(\Sigma^-$$\rightarrow$ n)=$+0.20\pm.08$.
The experimental measurement of $g_2$ is $g_2=-0.56\pm .37$.
Using the experimental value  $G_{A/V}(\Sigma^-$$\rightarrow$ n)=$+0.20\pm .08$
in the overall fit of the magnetic moment and beta decay ratios reduces $\chi^2$ to 16, without much change in any of the predictions.  This improvement in $\chi^2$ by relaxing the asumption that $g_2=0$ gives us more confidence in the model, and less in the assumption that $g_2=0$.

Ref.\ \cite{hsueh} was the only measurement of $G_A/G_V$ that allowed $g_2$ to vary in fitting the experimental distributions.  
The prediction that $g_2=0$ for neutron decay is based on isospin, and therefor is probably a safe conclusion. The other measurements (for $\Lambda$, $\Xi^-$, and $\Xi^0$) should be considered theoretical results, based on experiment and the assumption of SU(3) symmetry.   
Until these experiments are analyzed without the SU(3) assumption or new experiments performed,
we have to use this data as the only data to fit to, but it must be regarded as tentative.
There are indications that the ratio $g_2/G_V$ is positive in the quark model for these hyperon decays\cite{dh}.   That would give a positive contribution to these $G_{A/V}$ ratios, tending to improve our model's fit to the data.

\section{Quark spin projections and antiquark content of the proton.}

The quark and pion wave functions can be used to calculate 
the quark spin projections $\Delta u$, 
$\Delta d$, and the total quark spin projection $\Sigma$. 
The spin projection for quark $q$ is defined by
\begin{equation}
\Delta q = <\sum_i\sigma^i_z>_q,
\end{equation}
where the sum is over only type $q$ quarks.  The total quark spin projection $\Sigma$ is given by the sum of $\sigma_z$ over all quarks
\begin{equation}
\Sigma = <\sum_i\sigma^i_z> = \Delta u + \Delta d.
\end{equation}
It follows from isotopic spin rotation that the quark spin projections are related to $G_{A/V}$ for the neutron by
\begin{equation}
G_{A/V}(n\rightarrow p) = \Delta u -\Delta d.
\label{bj}
\end{equation}

It has to be emphasized here that these quark spin projections are for the proton in its rest system.
They are not the same as corresponding quark spin projections on the light cone at infinite momentum, which are calculated using QCD sum rules for polarized deep inelastic scattering asymmetries.  Since QCD is a strong interaction, a boost to infinite momentum produces gluons and quark-antiquark pairs that were not in the rest frame wave function.  This changes the individual and total quark spin projections.  Equation (\ref{bj}) is not affected by the boost if it is assumed that the quark pairs produced in the boost are charge symmetric.  It then becomes the well known Bjorken sum rule.

We find for the rest frame spin projections
\begin{equation}
\Delta u=0.98\pm .05,\quad \Delta d=-0.35\pm .01,\quad \Sigma=0.63\pm .06.
\end{equation}
While this shows a considerable decrease in total quark spin projection from the static quark model value $\Sigma=1$, it is not as great a decrease as that indicated  in QCD sum rules.
Note that, since this model has no SU(3) symmetry, $\Delta s=0$.

The pion component of the proton can be considered as a quark-antiquark sea in the rest frame wave function.  This has no quark spin projection because the pions are spin zero, but does contribute orbital angular momentum to the total angular momentum of the proton.  
The $z$ component is calculated 
as the expectation value of $L_z$ for the pions 
\begin{equation}
L_z=0.19\mp.06.
\end{equation}
As required by conservation of angular momentum, we see that
\begin{equation}
\frac{1}{2}\Sigma + L_z = \frac{1}{2}.
\end{equation}

Considering the pions as quark-antiquark pairs,
we can also calculate the antiquark content $\overline{u}$ and $\overline{d}$
of the proton.  We find
\begin{equation}
\overline{u}=0.07,\quad \overline{d}=0.26,\quad \overline{d}-\overline{u}=0.19\pm .03.
\end{equation}
With this value for $\overline{d}-\overline{u}$, the quark and antiquark contribution to the Gottfried sum rule\cite{ff} is
\begin{equation}
 S_G = \frac{1}{3}[1-2(\overline{d}-\overline{u})] =0.21\pm .02,
\end{equation}
in good agreement with the experimental result\cite{nmc} of $S_G=0.24\pm .03$.  This result would survive a boost because the quark pairs produced by QCD are expected to have equal numbers of 
$u$-$\overline u$ and $d$-$\overline d$ pairs. 

\section{Conclusion}

Our main conclusion is that a relatively simple phenomenological quark model
can provide a combined fit to the beta decay ratios and magnetic moments.  
The longstanding problem of reducing the static quark model prediction of $5/3$ for the neutron $G_{A/V}$ can be solved if there is a sizeable pion component in the nucleon, along with some orbital and decuplet excitation.  The pions in the proton wave function also provide the appropriate difference of 
$\overline{d}-\overline{u}$ antiquarks to satisfy the Gottfried sum rule.


\begin{thebibliography}{8}
\bibitem{jfpi} J. Franklin, Phys. Rev. D {\bf 30}, 1542 (1984).
This paper is referred to as I.  There are two misprints in I.
Equation (7) should not contain the constant $\gamma$
which is already included in the definition of the pion emission operator 
$\Theta$.  The right hand side of Eq.\ (14) should not be divided 
by nine.  Neither of these misprints affected the results because the
correct equations were used in all calculations.
\bibitem{sl}The sign convention for the transition moment $\mu_{\Sigma\Lambda}$ has been changed in this paper so that the quark model transition moment is now positive.  Consequently, there are some sign changes in the new equations.  There were also two errors in the d-quark contribution to 
$\mu_{\Sigma^+}$ and $\mu_{\Sigma^-}$ from decuplet states in I that have been corrected here.
These errors were small and had negligible effect on the results of I.
\bibitem{rcp} If this state is put into the wave function for the $\chi^2$ fit in Section 5, it turns out to have a 3\% probability, while lowering $\chi^2$ by less than one, so we have not bothered to include it in the fit.
\end{thebibliography}
\end{document}